\documentclass{appolbnh}
\usepackage{graphicx}
\usepackage{color}
\usepackage{lineno}
\usepackage{ulem}
\newcommand{\rr[1]}{{\color{red}{#1}}}

\begin{document}

\title{Anatomy of Roper Resonance %
}

\author{Igor Strakovsky
\thanks{igor@gwu.edu}
\address{Institute for Nuclear Studies, Department of Physics, \\
         The George Washington University, Washington, DC 20052, USA}
}

\maketitle

\begin{abstract}
Sixty years ago, the first excited state of a proton/neutron was ``born.''
During this time, we learned a lot about it\rr[,] specifically — how unique this 
case is: a single resonance with two pole positions on different Riemann 
sheets. Let me present a brief history to remind readers how development 
progressed. Sure, history is sometimes something that never happened, described by 
those who were never there...
\end{abstract}


\section{Introduction}
\begin{figure}[htb!]
\centering
{
    \includegraphics[width=0.55\textwidth,keepaspectratio]{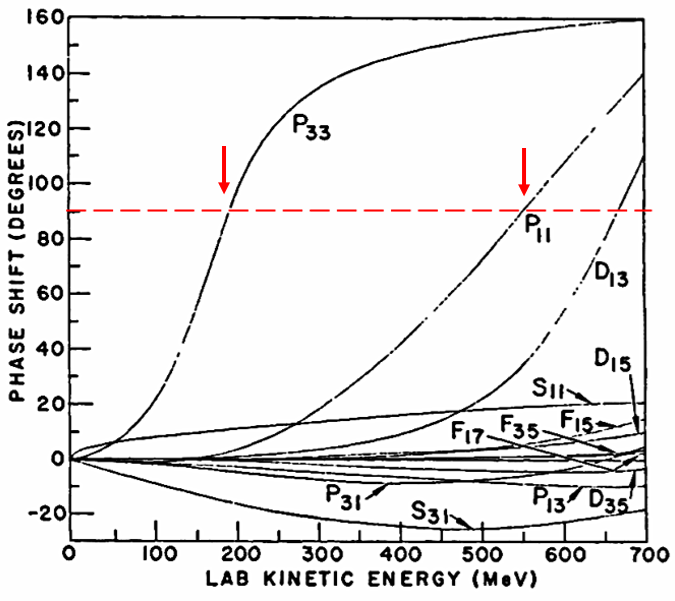} 
}

\centerline{\parbox{1\textwidth}{
\caption[] {\protect\small
Pion-nucleon phase shifts as functions of energy from 0 to $700~\mathrm{MeV}$~\cite{Roper:1964zza}. 
The resonance signal is associated with a phase crossing of $90^\circ$, as shown by the red vertical arrows. 
The previous discovery of the $\Delta(1232)3/2^+$ in $P_{33}$ was done by Fermi's group at the Chicago Synchrocyclotron~\cite{Anderson:1953dgc}, while now Dave Roper reported his observation for the $P_{11}$ case~\cite{Roper:1964zza}.}
\label{fig-1} } }
\end{figure}

\begin{figure}[htb!]
\centering
{
    \includegraphics[width=0.6\textwidth,keepaspectratio]{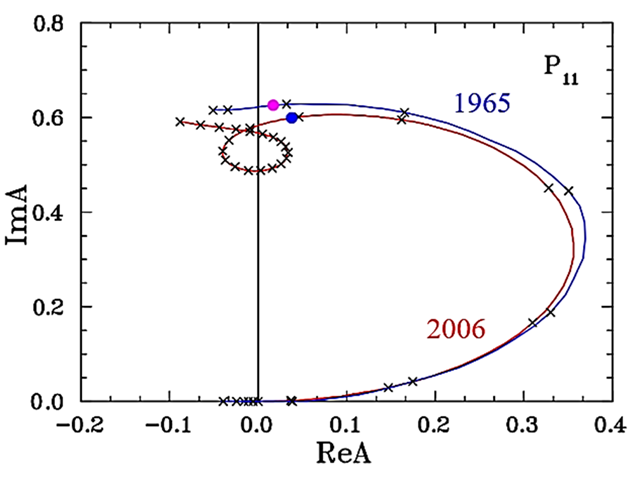} 
}

\centerline{\parbox{1\textwidth}{
\caption[] {\protect\small
Argand plots for partial-wave $P_{11}$ amplitude from threshold ($1080~\mathrm{MeV}$) to $W = 2.5~\mathrm{GeV}$. Blue (red) solid curve corresponds to the original Roper's amplitude~\cite{Roper:2025} (SAID SP06 amplitude~\cite{Arndt:2006bf}). Crosses indicate $50~\mathrm{MeV}$ steps in $W$.
Filled circles correspond to the Breit-Wigner (BW) $W_R$ determination.}
\label{fig-2} } }
\end{figure}

QCD gives rise to the hadron spectrum~\cite{Gell-Mann:1964ewy, Zweig:1964jf} and many $q\bar{q}$ and $qqq$ states have been observed~\cite{ParticleDataGroup:2024cfk}. Sixty years ago, the first excited state of a proton/neutron was discovered by Dave Roper in his PhD work at Massachusetts Institute of Technology
(MIT) and Livermore Radiation Laboratory (LRL). This was reported in 1963 at the Siena conference in Italy by Dave Roper and his PhD adviser, Bernard Feld~\cite{Roper:1963}. Then, at Michael Moravcsik's request, Bernard Feld agreed that Dave Roper's name should be the only author of this Physical Review Letters paper~\cite{Roper:1964zza}. That is how $N(1440)~1/2^+$ resonance was ``born'' (Fig.~\ref{fig-1}), and as a result, this resonance is widely known informally as the ``Roper'' resonance~\cite{Roper:2025}. During this time, the elastic $\pi^\pm p$ scattering database increased dramatically, but the situation with $N(1440)~1/2^+$ appears stable over 50~years, as Fig.~\ref{fig-2} and 
Table~\ref{tab-1} demonstrate.

\begin{figure}[htb!]
\centering
{
    \includegraphics[width=0.7\textwidth,keepaspectratio]{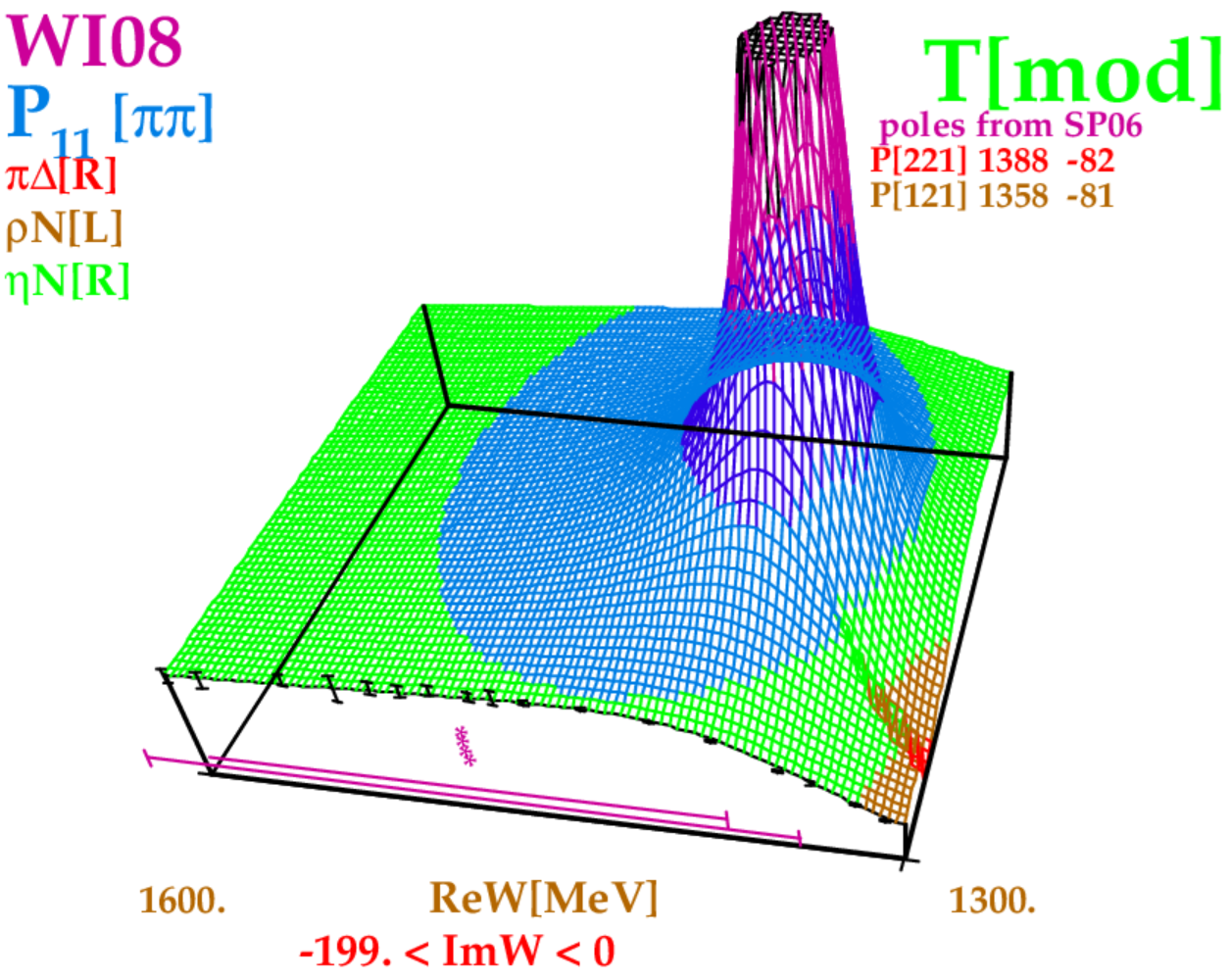} 
    \includegraphics[width=0.7\textwidth,keepaspectratio]{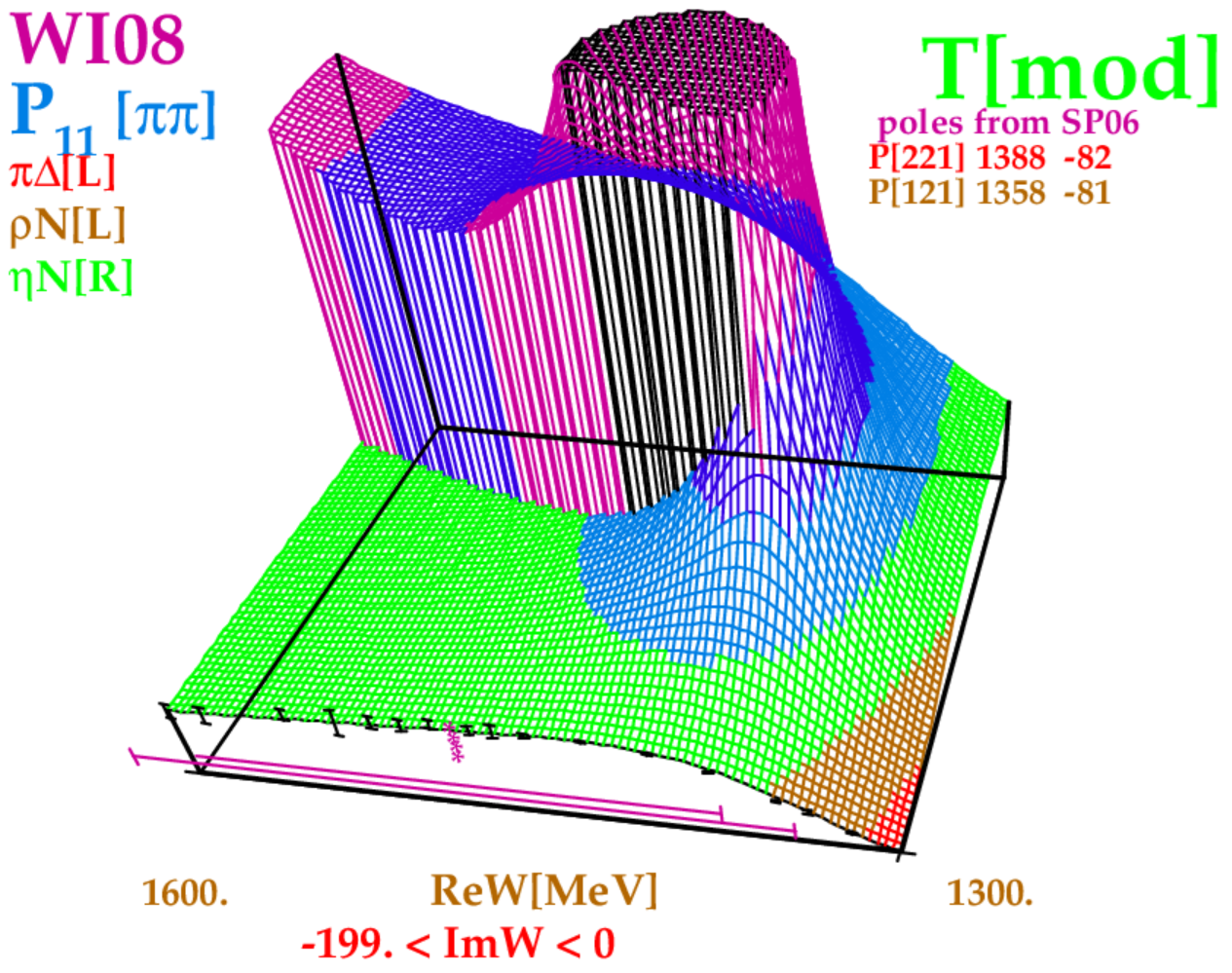} 
}

\centerline{\parbox{1\textwidth}{
\caption[] {\protect\small
Two poles for $\pi N$ $P_{11}$ for SAID WI08 amplitude~\cite{Workman:2012hx}. \underline{Top}: the $\pi\Delta$-cut can be seen in the foreground and runs from larger to smaller values of the real part of the energy. \underline{Bottom}: the $\pi\Delta$ cut is clearly visible running from smaller to larger values of the real part of the energy.} 
\label{fig-3} } }
\end{figure}

Discovered 60 years ago, the Roper resonance state has remained controversial ever since. The prominent $N(1440)~P_{11}$ resonance is clearly
evident in both Karlsruhe-Helsinki (KH) and The George Washington/Virginia Tech (GW/VPI) analyses (Figs.~4—7 from 
Ref.~\cite{Arndt:2006bf}), but it occurs very near the $\pi\Delta$ ($W = 1350-i~50~\mathrm{MeV}$), $\eta N$ ($W = 1487-i~0~\mathrm{MeV}$), and $\rho N$ ($W = 1715-i~73~\mathrm{MeV}$) thresholds (Fig.~8 from Ref.~\cite{Arndt:2003if}), making a BW fit questionable. The $N(1440)$ is unique in that its behavior on the real energy axis is influenced by poles on different Riemann sheets (with respect to the $\pi\Delta$-cut), as was first reported by Arndt \textit{et al.}~\cite{Arndt:1985vj}. This happened 20 years after the Roper resonance was 
born. Due to the nearby $\pi\Delta$ threshold, both $P_{11}$ poles are not far from the physical region (Fig.~\ref{fig-3}). There is a small shift between pole positions on the two sheets due to a non-zero jump at the $\pi\Delta$-cut. The conclusion is that a simple BW parametrization cannot account for such a complicated structure. This point was also emphasized by H\"ohler~\cite{Hohler:2008}.

\section{Two Pole Observation}
Resonances are defined by the poles of the $S$-matrix, whether in scattering, production, or decay matrix elements. Recent studies of the $N(1440)$ $1/2^+$ resonance by the J\"ulich~\cite{Doring:2009yv} and ANL-Osaka (EBAC)~\cite{Kamano:2010ud} groups have confirmed the two-pole determination. An earlier study by Cutkosky and Wang came to a similar conclusion~\cite{Cutkosky:1990zh}. Table~\ref{tab-1} summarizes the results of the phenomenological research conducted by several groups.  One can see that the results from different treatments are in reasonable agreement.

\begin{table}[htb!]
\centering \protect\caption{{Pole positions from the different phenomenological studies of the $N(1440)1/2^+$ resonance. Real($W_R$) and imaginary($-2W_I$) parts are listed.}
}
\vspace{4mm}
{%
\begin{tabular}{|c|c|c|}
\hline
Pole-1       & Pole-2     & Reference \tabularnewline
(MeV)        &  (MeV)     &  \tabularnewline
\hline
1359-i~100   & 1410-i~80  & \cite{Arndt:1985vj} \tabularnewline
1358-i~80    & 1388-i~82  & \cite{Workman:2012hx} \tabularnewline
1370-i~114   & 1360-i~120 & \cite{Cutkosky:1990zh} \tabularnewline
1387-i~73    & 1397-i~71  & \cite{Doring:2009yv} \tabularnewline
1357-i~76    & 1364-i~105 & \cite{Kamano:2010ud} \tabularnewline
\hline
\end{tabular}} \label{tab-1}
\end{table}

\section{Conclusion}
Overall, most of the analyses of $N(1440)$ are based on its BW parameterization, which implicitly assumes that the resonance is related to an isolated pole on the second Riemann sheet. However, 
given the complicated structure found in our SAID PWA, the BW description may only 
be an effective parameterization that could differ across various
processes. Some inelastic data indirectly support this point, giving $N(1440)$ BW 
masses and widths significantly different from the PDG BW 
values~\cite{Hohler:2008}. This may also cast some doubt on the recent $Q^2$ evaluation 
results~\cite{Drechsel:2007if, CLAS:2008roe, CLAS:2012wxw, Mokeev:2015lda, 
Mokeev:2023zhq}, since the $Q^2$-dependence of the contributions from different 
singularities may vary. This may also suggest the need to extend the extraction of 
$N(1440)~1/2^+$ electroexcitation amplitudes beyond the BW approximation to shed light on $Q^2$, the evolution of the two residues at the two pole positions. This problem can be studied in experiments at JLab CLAS12. Combining anticipated results from CLAS6 and CLAS12 will address an important open question: whether the two Roper poles originated from a common quark core. The important step in this direction was made in Ref.~\cite{Wang:2024byt, Suzuki:2010yn}.

More details about the Roper resonance are available in this volume of Acta Physica Polonica~B.

Finally, let me cite some words from well-known experts that I learned from: 
Dave Roper: \textit{I spent a lot of time trying to eliminate the $P_{11}$ resonance}, and Dick Arndt: \textit{This is one of the mysterious resonances}.

\section*{Acknowledgment}
I thank Dave Roper, Ron Workman, Michal Praszalowicz, and Victor Mokeev for their valuable comments and discussions.
This work was supported in part by the U.~S.~Department of Energy, Office of Science, Office of Nuclear Physics, under Award No.~DE--SC0016583.


\end{document}